\documentclass[a4paper]{jpconf}
\usepackage{graphics,graphicx}
\usepackage{amsmath,amsfonts,amssymb,amsthm,commath}
\usepackage{bm,times,dcolumn,dsfont}
\usepackage{float}
\usepackage[bookmarks=true]{hyperref}
\usepackage{cleveref}

\usepackage{cite}

\begin{document}

\title{Quantum Correlations in Jahn-Teller Molecular Systems Simulated with Superconducting Circuits}

\author{Ali Pedram, Onur Pusuluk, and \"{O} E. M\"{u}stecapl{\i}o\u{g}lu}

\address{Department of Physics, Ko\c{c} University, Sar{\i}yer, Istanbul, 34450, Turkey}

\ead{apedram19@ku.edu.tr,opusuluk@ku.edu.tr,omustecap@ku.edu.tr}

\begin{abstract}
We explore quantum correlations, in particular, quantum entanglement, among vibrational phonon modes as well as between electronic and vibrational degrees of freedom in molecular systems, described by Jahn-Teller mechanism. Specifically, to isolate and simplify the phonon- electron interactions in a complex molecular system, the basis of our discussions is taken to be the proposal of simulating two-frequency Jahn- Teller systems using superconducting circuit quantum electrodynamics systems (circuit QED) by Tekin Dereli and co-workers in 2012. We evaluate the quantum correlations, in particular entanglement between the vibrational phonon modes, and present analytical explanations using a single privileged Jahn-Teller mode picture. Furthermore, spin-orbit entanglement or quantum correlations between electronic and vibrational degrees of freedom are examined, too. We conclude by discussing experimental feasibility to detect such quantum correlations, considering the dephasing and decoherence in state-of-the-art superconducting two-level systems (qubits).
\end{abstract}

\section{Introduction}
Quantum entanglement in natural systems, in particular molecules, has been attracted much attention for its fundamental and practical significance~\cite{boguslawski_orbital_2013,pusuluk_classical_2021,ding_concept_2021,kurashige_entangled_2013,hou_entanglement_2005,hines_entanglement_2004,hou_entanglement_2006,liberti_entanglement_2007,streltsov_jahn-teller_2020,boguslawski_orbital_2015,lin_quantum_2020,mckemmish_quantum_2011,mottet_quantum_2014-1,barcza_quantum-information_2011,szalay_correlation_2017}. Due to their complex structure and many interactions within, it is difficult to isolate the quantum entanglement route in molecules. One such route is the case where electronic and vibrational degrees of freedom are coupled via so-called Jahn-Teller interaction~\cite{mckemmish_quantum_2011,streltsov_jahn-teller_2020,hines_entanglement_2004,liberti_entanglement_2007}. Dereli and coworkers have proposed that quantum simulation of molecules using superconducting circuits can serve to capture the essential physics of multi-mode Jahn-Teller effect~\cite{dereli_two-frequency_2012}. Here, we take advantage of the Dereli and co-workers’ methods and explore quantum entanglement in two-mode Jahn-Teller molecules simulated by superconducting circuits .

The contribution by Dereli and coworkers is not only serving for the purpose of illuminating complex quantum processes in molecules but also provides an unprecedented application direction for the ultrastrong coupling regime of superconducting circuit QED. Jahn-Teller coupling can be comparable to involved frequencies of the vibrational phonons and hence it requires an artificial system where ultrastrong coupling can be realized~\cite{hines_entanglement_2004,larson_jahn-teller_2008,meaney_jahn-teller_2010}. Realization of such strong interaction regimes are challenging in the usual cavity QED but it is possible with the advent of superconducting circuit QED~\cite{yoshihara_superconducting_2017,forn-diaz_ultrastrong_2019,frisk_kockum_ultrastrong_2019}. Flux qubits and transmission line resonators are ideal platforms to realize such strong photon-qubit interactions, where photons can play the role of vibrational phonons. Vibrational modes in Jahn-Teller molecules can be optical and acoustic phonons and hence an ideal platform to simulate such a system would need tunable and wide range of frequencies in addition to the strong coupling. In the case of superconducting circuits even optomechanical-like interactions can be simulated by using transmission line resonators~\cite{johansson_optomechanical-like_2014,eichler_realizing_2018,rimberg_cavity-cooper_2014,kim_circuit_2015}.

Typical focus on Jahn-Teller molecules in chemistry is limited to energetically low-lying states and their spectral properties, such as optical absorption. We explore quantum correlations and look for potential quantum entanglement in ground state of the molecular system with two-mode Jahn-Teller coupling. Due to routine methods of measuring quantum correlations in superconducting circuit QED, our results can be significant for practical entanglement generation in addition to illuminating the nature of entanglement in natural molecules. In addition, our quantum simulation and effective privileged mode approach can be further extended to explore the interplay of quantum correlations, chaos, and phase transitions~\cite{majernikova_quantum_2011}.

The rest of the paper is organized is as follows. We first give a short review of the two-mode Jahn-Teller model and the method of privileged mode as described by Dereli and coworkers in Sec. II. We present our quantum entanglement measures and their evaluation results in Sec. III. Sec. IV is the conclusion of our analysis of the quantum entanglement results in different parameter regimes from the perspective of effective privileged mode approach.

\section{Privileged mode approach to two-mode Jahn-Teller model: A short review}

\subsection{two-mode Jahn-Teller model and circuit QED simulation}

Jahn-Teller transition happens in some molecules that can end in a preferred geometrical configuration among a set of degenerate ones due to the symmetry breaking by vibrational phonon and localized electron interaction. Hamiltonian describing this phenomenon is known as Jahn-Teller model~\cite{bersuker_jahn-teller_2006}.  Jahn-Teller coupling belongs to the general class of spin-boson interactions. Vibrational degrees of freedom of molecules are quantized in terms of phonons with boson statistics. Optical and acoustic phonons have high and low frequencies and can be described by an approximate two-mode model, a particular Jahn-Teller model, also known as the Herberg-Teller model, which will be our focus here. Dereli and coworkers have shown that E$\times (\beta_1+\beta_2)$ type Herzberg-Teller model can be simulated by using the superconducting circuit (sc) Hamiltonian (we take $\hbar=1$)
\begin{eqnarray}
	\hat{H}_{\text{sc}} = \hat{H}_{0}+\hat{H}_{\text{int}}
\end{eqnarray}
where the non-interacting and interacting terms respectively are given by
\begin{eqnarray}
	 \hat{H}_{0} &=&	\frac{\omega_q}{2}\hat{\sigma}_z+\sum_{i=1}^2\omega_i
	 \hat{a}_i^\dag \hat{a}_i,\\
	 \hat{H}_{\text{int}} &=& \sum_{i=1}^2 g_i(\hat{a}_i+\hat{a}_i^\dag)\hat{\sigma}_x+J( \hat{a}_1^\dag \hat{a}_2+ \hat{a}_2^\dag \hat{a}_1).
\end{eqnarray}
Here, $\omega_q$ and $\omega_i$ denote the transition frequency of the qubit and the frequencies of the modes of two lumped element LC resonators  ($i=1,2$). Creation ($\hat{a}_i^\dag$) and annihilation ($\hat{a}_i$) operators of the modes obey the Weyl-Heisenberg (bosonic) algebra. The two-qubit states simulate the diabatic electronic states in actual molecules. Two energy levels of the qubit, $|1\rangle,|2\rangle$ are represented by the Pauli spin-$1/2$ operators $\hat{\sigma}_z=|2\rangle\langle 2|-|1\rangle\langle 1|,\hat{\sigma}_x=|1\rangle\langle 2|+|2\rangle\langle 1|$. The resonators are coupled to a flux qubit with the interaction coefficients $g_i$. Their difference can simulate the case of spatial anisotropy of molecules. We scale the interaction coefficients with the corresponding mode frequencies and write $g_i=\omega_ik_i$. The displacement-spin coupling form of the interaction is the signature of the Jahn-Teller model. The last term, however, makes the circuit QED model different than the original Jahn-Teller Hamiltonian. It arises due to the inductive coupling of the two resonators. This term can be made relatively smaller than the ultrastrong coupling coefficient by optimizing the system geometry by extending the distance between the LC resonators using a sufficiently large flux qubit. In the privileged mode description, this term will not play a significant role.

\subsection{Privileged Mode Approach to Two-Mode Jahn-Teller Model}

Privileged mode is an energetically favored superposition of the two resonator modes which can be expressed in terms of the annihilation operators as
\begin{eqnarray}
\hat{b}_1=\frac{1}{k_p}(k_1\hat{a}_1+k_2\hat{a}_2),\\
\hat{b}_2=\frac{1}{k_p}(k_2\hat{a}_1-k_1\hat{a}_2),
\end{eqnarray}
where $k_p^2=k_1^2+k_2^2$. Next to the privileged mode, there is also an energetically disadvantaged mode, whose annihilation operator is denoted by $\hat{b}_2$, and the transformation from the resonator modes to the privileged and disfavored modes is one-to-one. The orthogonal transformation coefficients are determined by maximizing $k_p^2\omega_p$ and using the orthogonality condition of the transformation~\cite{barentzen_canonical_1978}.

The terms of the transformed Hamiltonian for the privileged mode is the same as the single mode Jahn-Teller model
\begin{eqnarray}
H_{\text{JT}}=\frac{\omega_q}{2}\hat{\sigma}_z+\omega_p
\hat{b}_1^\dag \hat{b}_1+g_p(\hat{b}_1+\hat{b}_1^\dag)\hat{\sigma}_x,
\end{eqnarray}
where the privileged mode frequency and its coupling rate to the flux qubit are denoted by $\omega_p=(\omega_1k_1^2+\omega_2k_2^2)/k_p^2$ and $g_p=\omega_p k_p$, respectively.
The rest of the total transformed Hamiltonian terms include the hopping between the privileged and disadvantaged modes. The full hamiltonian for the superconducting circuit after the transformation becomes
\begin{eqnarray} \label{Eq::TransformedHam}
    \hat{H}_{\text{sc}} = \frac{\omega_q}{2}\hat{\sigma}_z + (\omega_p+\frac{Jk_1k_2}{k_p^2})\hat{b}_1^\dag \hat{b}_1 + (\tilde{\omega}_p-\frac{Jk_1k_2}{k_p^2})\hat{b}_2^\dag \hat{b}_2 + [c+J\frac{k_2^2-k_1^2}{k_p^2}](\hat{b}_1^\dag \hat{b}_2+\hat{b}_2^\dag \hat{b}_1) \\
  +k_p[\omega_p(\hat{b}_1^\dag+\hat{b}_1)+c(\hat{b}_2^\dag+\hat{b}_2)]\hat{\sigma}_x. \nonumber
\end{eqnarray}
Here we have $\tilde{\omega}_p = (\omega_1k_2^2+\omega_2k_1^2)/k_p^2$. To ensure the terms for the disadvantaged mode are negligible, perturbation theory requires that the parameters $J$ and $c=\Delta k_1k_2/k_p^2$ must be sufficiently small where $\Delta = \omega_1-\omega_2$. The latter condition can be realized if one resonator mode interacts with the flux qubit in the ultrastrong coupling regime (without loss of generality, we label it as $i=1$). The other mode, $i=2$, can be coupled strongly to the flux qubit, and we demand the condition $g_1>g_2$ to be satisfied. The hopping rate $J$ between the modes must be smaller or comparable to $g_2$.

\section{Results: Quantum Entanglement in the Jahn-Teller Ground State and Privileged Mode Interpretation}

We quantify the quantum entanglement in the ground state of the Jahn-Teller model using the logarithmic negativity~\cite{PhysRevLett.95.090503} as the measure of entanglement. The logarithmic negativity, which is an upper bound to the distillable entanglement, is defined as
\begin{eqnarray}
E_N(A|\bar{A})=log_2\norm{\rho_{A\bar{A}}^{T_A}},
\end{eqnarray}
where $A$ is a subsystem of a total quantum system in which its complement is $\bar{A}$, $\rho^{T_A}$ signifies partial transpose with respect to the subsystem $A$, and $\norm{\rho}$ is the trace norm that equals to $\mathrm{tr}[\sqrt{\rho \rho^\dagger}]$. Here, $E_N(A|\bar{A})$ quantifies the entanglement shared between the subsystems $A$ and $\bar{A}$ when their joint state is described by the density matrix $\rho_{A\bar{A}}$.

For the case of our model, correlations might arise between the flux qubit ($S$), the privileged mode ($B_1$), and the disadvantaged mode ($B_2$). To capture an adequate description of the correlations in the tripartite ground state $\rho_{SB_1B_2}$ due to the Jahn-Teller Hamiltonian, we consider the bipartitions $S|B_1B_2$, $S|B_1$, $S|B_2$, and $B_1|B_2$ in what follows. To measure the entanglement shared between two of the three subsystems, we first compute the reduced density matrix of the subsystems under consideration by tracing out the degrees of the freedom of the third one.

According to our calculations of $E_N(S|B_1B_2)$, $E_N(S|B_1)$, $E_N(S|B_2)$, and $E_N(B_1|B_2)$, it is sufficient to set the Fock space dimension $N$ of the resonators to $10$ to analyze the behaviour of ground-state quantum entanglement with respect to the Hamiltonian parameters. We have done the same calculations for higher-dimensional Hilbert spaces up until $N =20$, and they gave similar results for the strong and ultra-strong coupling regimes.
\begin{figure}[t]
  \centering
  \includegraphics[width=12cm]{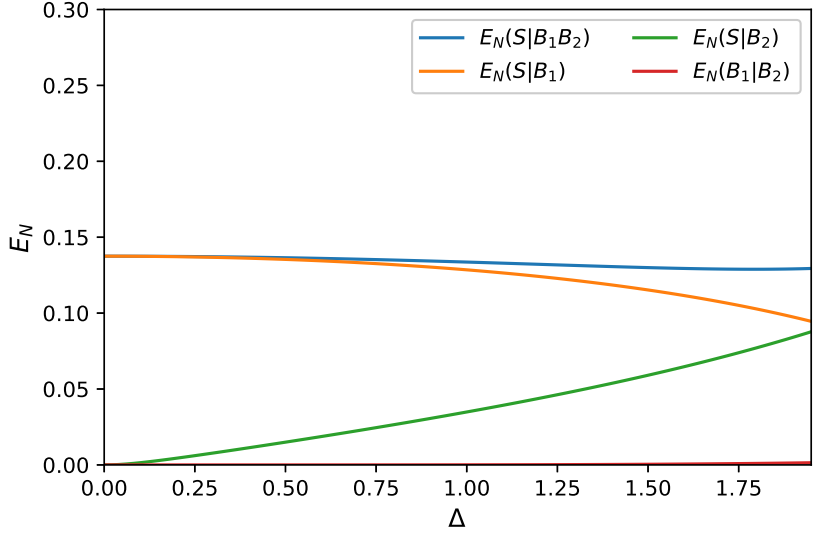}
  \caption{Ground state entanglement in Jahn-Teller model in terms of the logarithmic negativity $E_N$. $\Delta = \omega_1 - \omega_2$ is the frequency difference between the vibration modes. The parameters are fixed to explore the strong coupling regime such that $J = 0$ and $k_1 = k_2 = k = 0.1/\sqrt{2}$.}\label{Fig::Delta4kEq0p1}
\end{figure}
\begin{figure}[h]
  \centering
  \includegraphics[width=12cm]{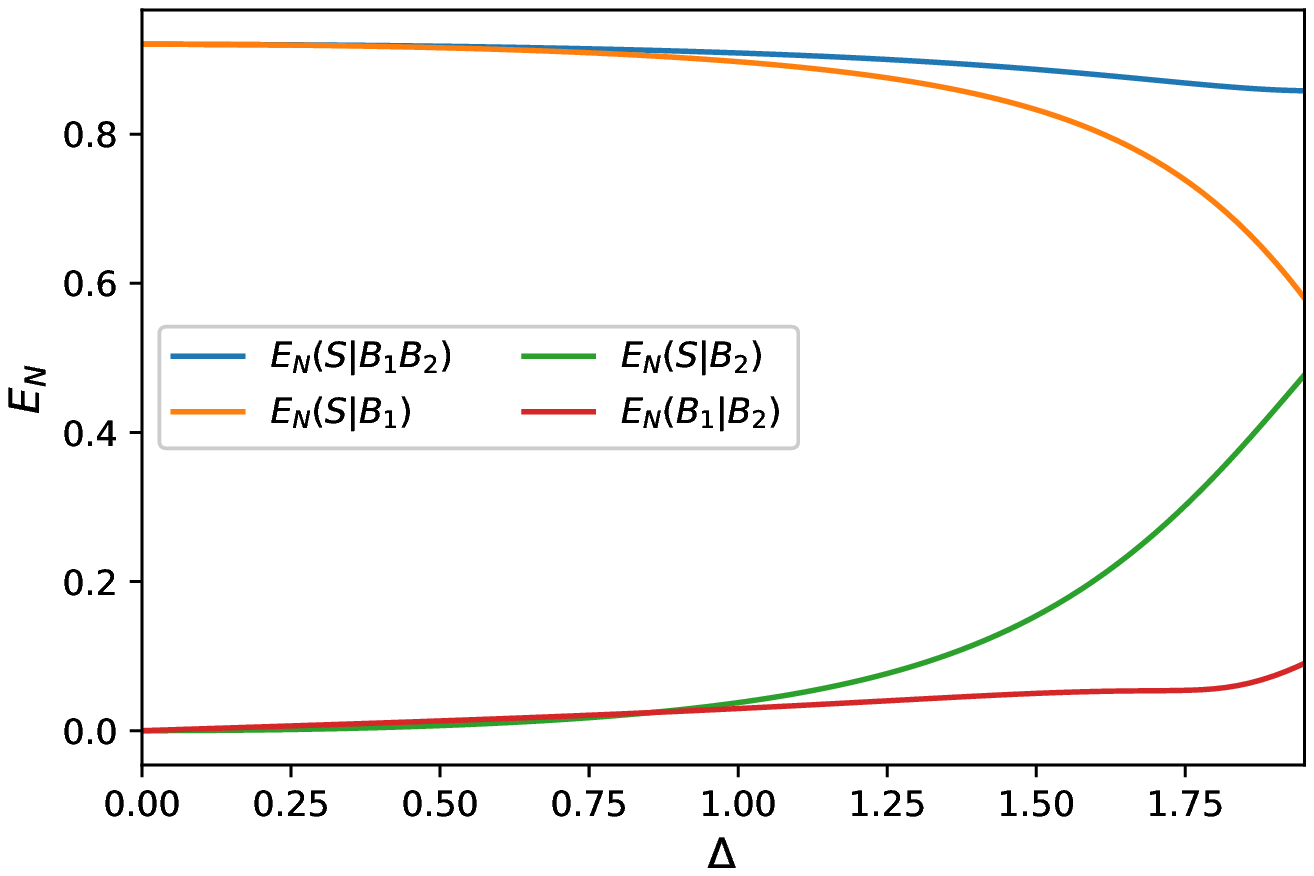}
  \caption{Ground state entanglement in Jahn-Teller model in terms of the logarithmic negativity $E_N$. $\Delta = \omega_1 - \omega_2$ is the frequency difference between the vibration modes. The parameters are fixed to explore the ultra-strong coupling regime such that $J = 0$ and $k_1 = k_2 = k = 1/\sqrt{2}$.}\label{Fig::Delta4kEq1}
\end{figure}
\begin{figure}[h]
  \centering
  \includegraphics[width=12cm]{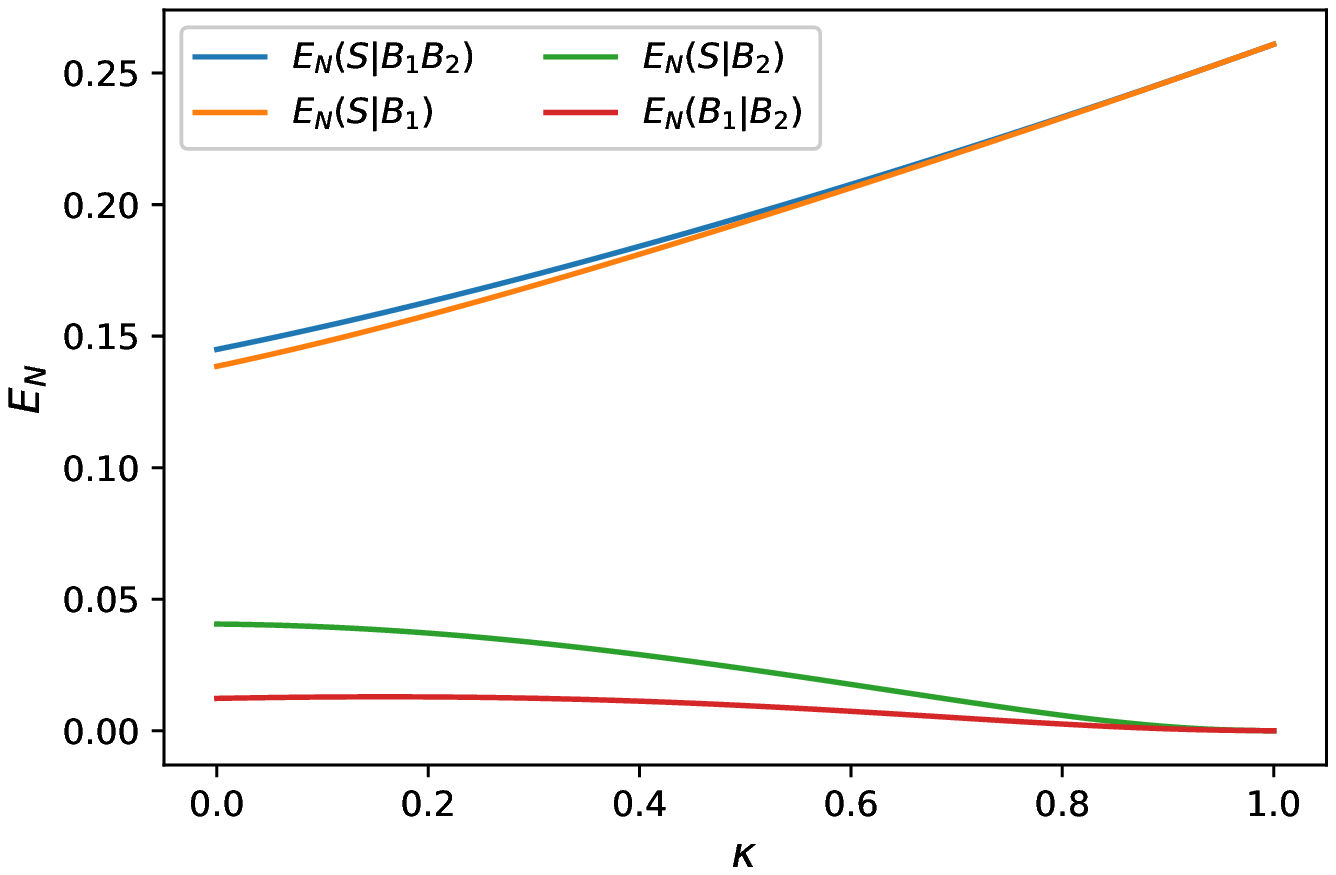}
  \caption{Ground state entanglement in Jahn-Teller model in terms of the logarithmic negativity $E_N$. $\kappa = k_2 - k_1$ is the difference between the factors that rescale the coupling constants $g_i$ with respect to frequencies $\omega_i$. The parameters are fixed such that $J = 0$ and $\omega_1 = 2 \omega_2 = 0.1$.}\label{Fig::k4DeltaEq0p05}
\end{figure}
\begin{figure}[h]
  \centering
  \includegraphics[width=12cm]{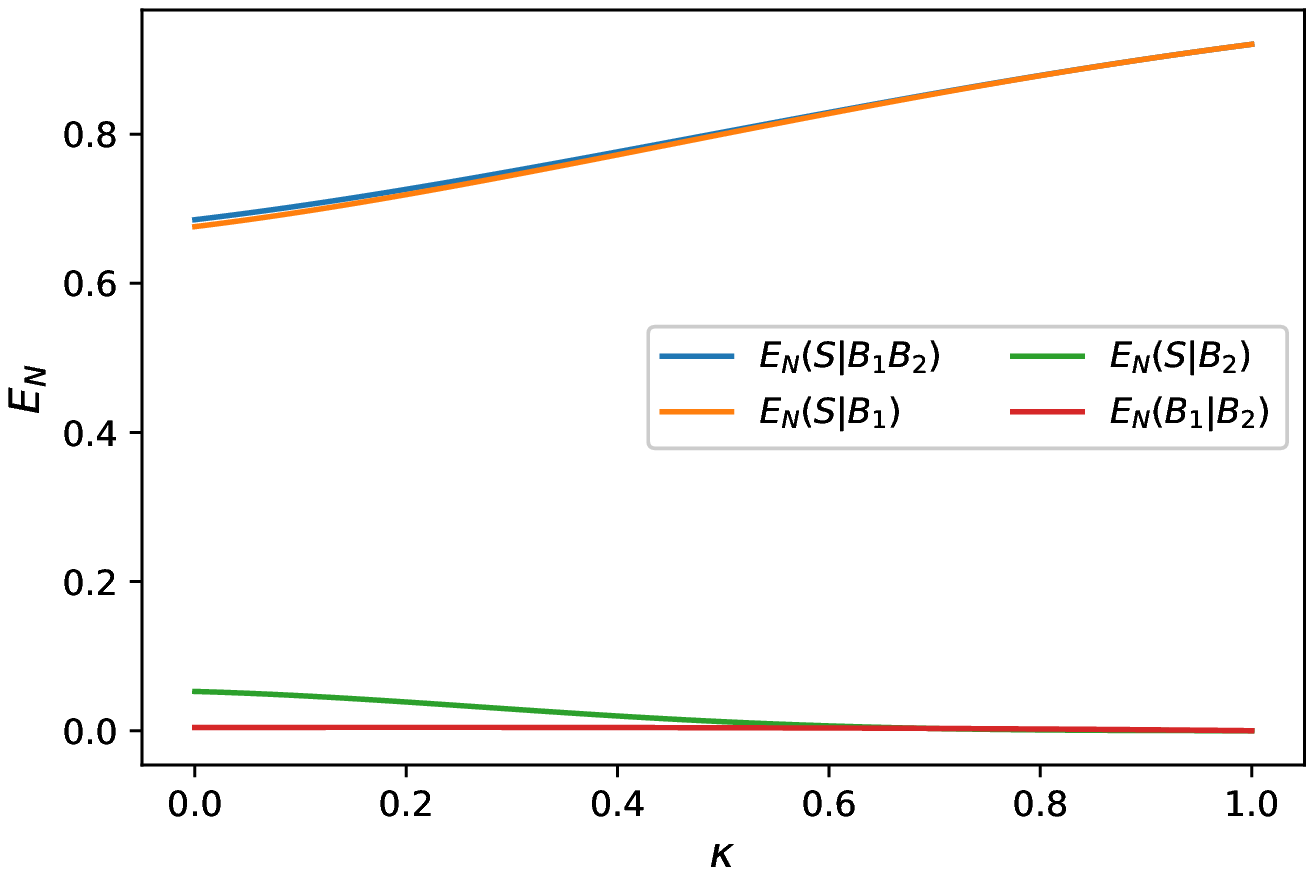}
  \caption{Ground state entanglement in Jahn-Teller model in terms of the logarithmic negativity $E_N$. $\kappa = k_2 - k_1$ is the difference between the factors that rescale the coupling constants $g_i$ with respect to frequencies $\omega_i$. The parameters are fixed such that $J = 0$ and $\omega_1 = 2 \omega_2 = 1$.}\label{Fig::k4DeltaEq0p5}
\end{figure}
\begin{figure}[h]
  \centering
  \includegraphics[width=12cm]{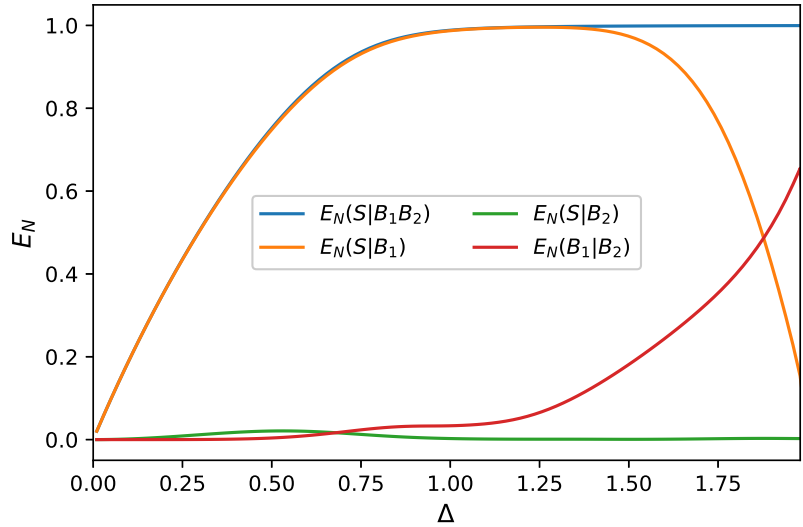}
  \caption{Ground state entanglement in Jahn-Teller model in terms of the logarithmic negativity $E_N$ when $J=0$ and $\omega_1 - \omega_2 = \Delta = k_1 = k_2$.}\label{Fig::kEqDelta}
\end{figure}
\begin{figure}[h]
  \centering
  \includegraphics[width=12cm]{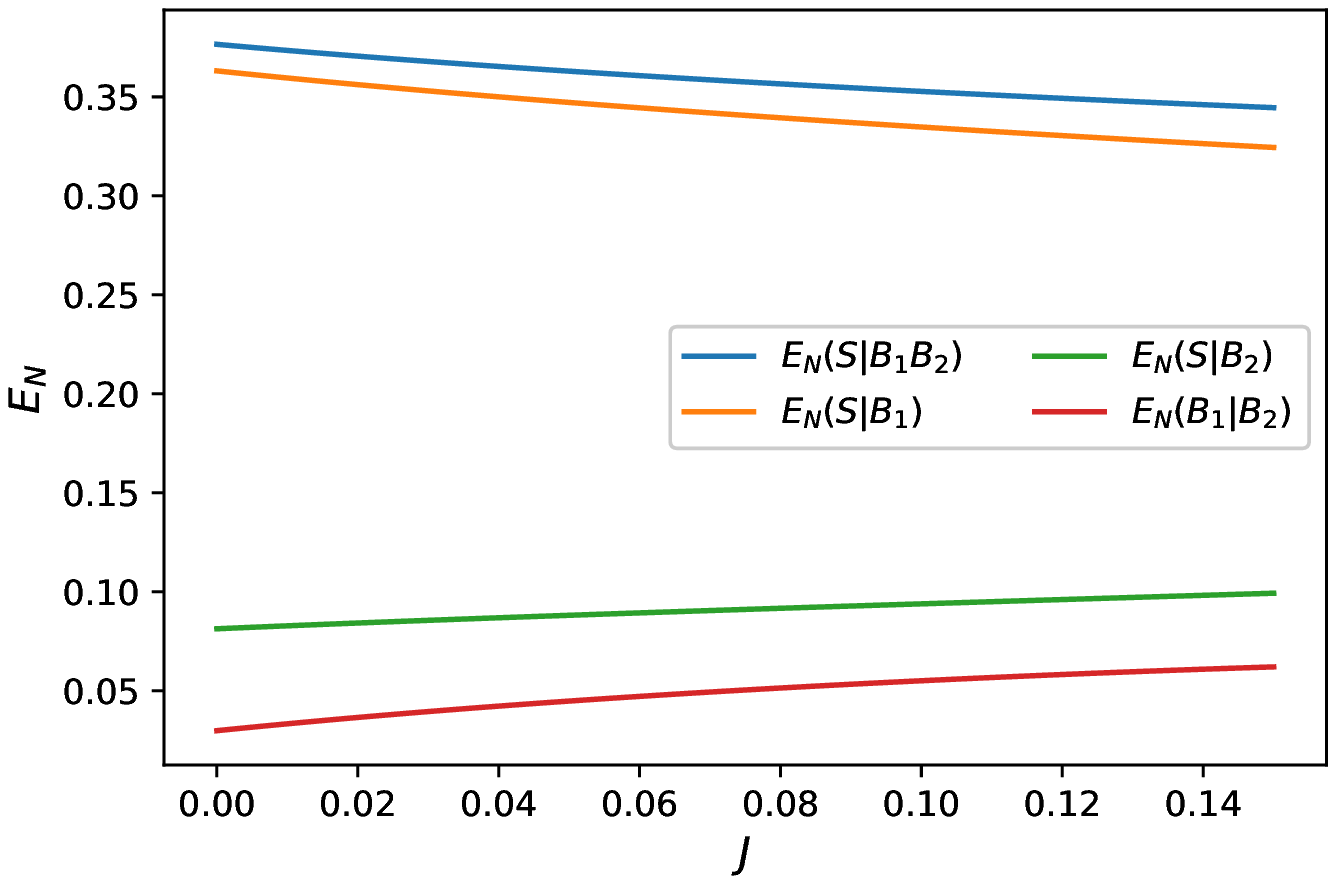}
  \caption{Ground state entanglement in Jahn-Teller model in terms of the logarithmic negativity $E_N$ when $J\neq0$. Here, $k_1 = k_2 = k =1/\sqrt{2}$ and $\omega_1 = 2 \omega_2 = 0.2$.}\label{Fig::J}
\end{figure}

We first examine the effect of the frequency difference between the vibration modes on the entanglement shared in the Jahn-Teller ground state. To this end, and for the sake of simplicity, we take $k_1 = k_2 = k$, $J = 0$ and $\omega_{1/2} = 1 \pm \Delta/2$ by rescaling all the parameters with respect to the qubit transition frequency $\omega_q$. It is clear from Eq.~\ref{Eq::TransformedHam} that as $\Delta$ is increased, both the coupling between the two resonators and the coupling of the flux qubit to the disadvantaged mode becomes stronger. Therefore, the flux qubit and the disadvantaged mode correlations are expected to constitute a significant portion of the total correlations as we increase $\Delta$. Fig.~\ref{Fig::Delta4kEq0p1} justifies this expectation in the strong coupling regime where $k = 0.1/\sqrt{2}$. The effective single privileged mode description of the system is valid for $|\Delta| < 0.1$. The entanglement between the flux qubit and the disadvantaged mode remains negligible until this limit. In the two-mode regime, $E_N(S|B_2)$ rises slowly with further increases of $\Delta$. When $\Delta = 2$, the flux qubit becomes coupled to both modes with the same strength, as a result of which $E_N(S|B_1)$ and $E_N(S|B_2)$ take the same value. Although the hopping constant $J$ is set to zero, $E_N(B_1|B_2)$ takes a small but nonzero value at this point.

For the case of ultra-strong coupling ($k=1/\sqrt{2}$) shown in Fig.~\ref{Fig::Delta4kEq1}, the general trend of ground-state entanglement does not appear different from the strong coupling case. However, the stronger coupling produces stronger entanglement between the flux qubit and the two resonators in this case. The observation of the steeper increase in the entanglement between the flux qubit and the disadvantaged mode is delayed until $|\Delta| = 1$, only after which the privileged mode approximation fails.

We also explore the ground state entanglement by changing the difference between the scaling factors appearing in the Jahn-Teller couplings $g_i = k_i \omega_i$. To do so, we set $J = 0$, $\omega_1 = 2 \omega_2$, and $k_{2/1} = (1 \pm \kappa)/2$. We use $k_2 - k_1 =\kappa$ as our control parameter. The results for $\omega_1 = 0.1$ and $\omega_1 = 1$ are respectively given in Figs.~\ref{Fig::k4DeltaEq0p05}~and~\ref{Fig::k4DeltaEq0p5}. When the flux qubit and the privileged mode are resonant, i.e., $\omega_q = \omega_1 = 1$, we end up with higher values for $E_N(S|B_1)$. The coupling of the privileged mode to the flux qubit is one order of magnitude larger than both the coupling of the disadvantaged mode and the hopping between the modes for $\kappa \geq 2/3$ in both cases. So, $E_N(S|B_1)$ approximately equals $E_N(S|B_1B_2)$ in both cases as long as the privileged mode approximation works well.

So far, a significant amount of entanglement has not appeared between the privileged and disadvantaged modes in the two-mode Jahn-Teller regime. This led us to change both $\omega_i$ and $k_i$ at the same time. For the sake of simplicity, we set $J = 0$, $\omega_{1/2} = 1 \pm \Delta/2$, and $k_1 = k_2 = \Delta$ in Fig.~\ref{Fig::kEqDelta}. Here, the privileged mode approximation fails when $\Delta \geq 0.2$. However, $E_N(S|B_1B_2)$ and $E_N(S|B_1)$ seem to overlap until $\Delta$ reaches 1. $E_N(B_1|B_2)$ rises rapidly after this point with further increases of $\Delta$. As of $\Delta = 1.5$, $E_N(S|B_1)$ shows a sharp decline, whereas $E_N(S|B_1B_2)$ remains fixed at unit negativity. When the flux qubit becomes coupled to both modes with the same strength at $\Delta = 2$, a significant amount of entanglement arises between the flux qubit and the two-mode subsystem and between the two modes themselves. However, neither the privileged nor the disadvantaged are strongly entangled with the flux qubit in this limit.  

Finally, we consider the role of hopping constant $J$ on the ground state entanglement in Fig.~\ref{Fig::J}. Here, the remaining parameters are fixed such that $k_1 = k_2 = k =1/\sqrt{2}$ and $\omega_1 = 2 \omega_2 = 0.2$, for which the privileged mode approximation works well for every value of $J$, i.e., the privileged mode is coupled to the flux qubit with a strength one order of magnitude larger than both its coupling strength to the disadvantaged mode and the flux qubit-disadvantaged mode coupling strength. However, $E_N(B_1|B_2)$ and $E_N(S|B_2)$ are always comparable to $E_N(S|B_1B_2)$ and $E_N(S|B_1)$. If we doubled $k$, $E_N(S|B_1B_2)$ would also double but $E_N(B_1|B_2)$ would increase more than six times in this figure. Thus, a non-vanishing hopping between the two modes results in nontrivial changes in the distribution of entanglement between different subsystems in the Jahn-Teller model.


\section{Conclusion}

We examined the quantum entanglement in the two-mode Jahn-Teller ground state simulated by a superconducting circuit using privileged mode description following the approach by Dereli and co-workers~\cite{dereli_two-frequency_2012}.

We find that the two-mode Jahn-Teller ground state exhibits bipartite entanglement between the qubit and both modes in the strong coupling regime of the superconducting circuit quantum simulator. This entanglement is weakly changed with the frequency difference (detuning) between the modes. The disadvantaged mode is approximately unentangled from the qubit at slight frequency differences, and it is unentangled from the privileged mode at all detunings. As the detuning increases towards the regime where the privileged mode approximation fails, both modes have comparable entanglement to the qubit. Similar behavior with stronger entanglement is found in the ultrastrong regime, too. The entanglement between the privileged mode and the qubit increases significantly, while the disadvantaged mode entanglement to the qubit is slowly decreasing, with the difference of the Jahn-Teller coupling constants, simulated by the flux qubit and the transmission line resonator interaction. We identified a parameter regime to entangle the privileged mode to the disadvantaged mode where both modes are equally strongly coupled to the qubit, and their detuning is significant so that the privileged mode approximation fails. In this case, the qubit is disentangled from the modes, and we only have entanglement between the bosonic modes. These conclusions are made when the modes are not directly coupled. We have also investigated the effect of direct coupling between the modes, which can be possible technically in superconducting circuits. Direct interaction of the modes can further increase the qubit-mode and the mode-mode entanglements. The enhancement is, however, more pronounced for mode-mode entanglement. 

Our results can be significant to simulate molecular systems with strong and ultrastrong coupled superconducting circuits to illuminate their quantum correlations and utilize them as large-scale quantum resources for quantum technologies.

\subsection{Acknowledgments}
O.~P.~ acknowledges support by the Scientific and Technological Research Council of Turkey (T\"{U}B\.{I}TAK), Grant No. (120F089).


\section*{References}
\bibliography{DereliPaperRefs}

\end{document}